\documentclass[numberedappendix]{emulateapj}

\newcommand{\himpc}{{\hbox {$~h^{-1}$}{\rm ~Mpc}}}
\newcommand{\hmpci}{{\hbox {$~h{\rm ~Mpc}^{-1}$}}}

\newcommand{\xirm}{\xi^{(r)}_m(r)}

\newcommand{\pkrm}{P^{(r)}_m(k)}
\newcommand{\jcap}{J. Cosmology Astropart. Phys.}

\newcommand{\tableskten}{&&&&\\[-6pt]}

\shorttitle{Determination of the Growth Factor from Redshift Distortions}
\shortauthors{Okumura \& Jing}

\begin{document}

\title{Systematic Effects on Determination of the Growth Factor \\ 
from Redshift-space Distortions}

\author{Teppei Okumura\altaffilmark{1,2} and Y.~P. Jing\altaffilmark{2}}

\email{teppei@ewha.ac.kr} 

\altaffiltext{1} {Institute for the Early Universe, Ewha Womans
  University, Seoul, 120-750, Korea} 

\altaffiltext{2} {Key Laboratory for Research in Galaxies and
  Cosmology, Shanghai Astronomical Observatory, Chinese Academy of
  Sciences, 80 Nandan RD, Shanghai, 200030, China}

\begin{abstract}
The linear growth factor of density perturbations is generally
believed to be a powerful observable quantity of future large redshift
surveys to probe physical properties of dark energy and to distinguish
among various gravity theories. We investigate systematic effects on
determination of the linear growth factor $f$ from a measurement of
redshift-space distortions. Using a large set of high-resolution
$N$-body simulations, we identify dark matter halos over a broad mass
range. We compute the power spectra and correlation functions for the
halos and then investigate how well the redshift distortion parameter
$\beta\equiv f/b$ can be reconstructed as a function of halo mass both
in Fourier and in configuration space, where $b$ is the bias
parameter.  We find that the $\beta$ value thus measured for a fixed
halo mass generally is a function of scale for $k>0.02\hmpci$ in
Fourier space or $r<80\himpc$ in configuration space, in contrast with
the common expectation that $\beta$ approaches a constant described by
Kaiser's formula on the large scales. The scale dependence depends on
the halo mass, being stronger for smaller halos. It is complex and
cannot be easily explained with the exponential distribution function
in configuration space or with the Lorentz function in Fourier space
of the halo peculiar velocities. We demonstrate that the biasing for
smaller halos has larger nonlinearity and stochasticity, thus the
linear bias assumption adopted in Kaiser's derivation become worse for
smaller halos. Only for massive halos with the bias parameter $b\geq
1.5$, the $\beta$ value approaches the constant predicted by the
linear theory on scales of $k<0.08\hmpci$ or $r>30\himpc$. Luminous
red galaxies (LRGs), targeted by the Sloan Digital Sky Survey (SDSS)
and the SDSS-III's Baryon Oscillation Spectroscopic Survey (BOSS),
tend to reside in very massive halos. Our results indicate that if the
central LRG sample is used for the measurement of redshift-space
distortions, fortunately the linear growth factor can be measured
unbiasedly.  On the other hand, emission line galaxies, targeted by
some future redshift surveys such as the BigBOSS survey, are inhabited
in halos of a broad mass range.  If one considers to use such
galaxies, the scale dependence of $\beta$ must be taken into account
carefully; otherwise one might give incorrect constraints on dark
energy or modified gravity theories. We also find that the $\beta$
reconstructed in Fourier space behaves fairly better than that in
configuration space when compared with the linear theory prediction.
\end{abstract}


\section{Introduction}
The presence of dark energy, which changes the gravitational assembly
history of matter in the universe, explains observed acceleration of
the cosmic expansion well within the framework of general relativity
\citep[see Komatsu et al. 2010 for the latest constraints]{Riess1998,
  Perlmutter1999, Spergel2003}.  There are also many attempts to
explain the acceleration without dark energy by modifying general
relativity on cosmological scales \citep[see, e.g.,][]{Dvali2000,
  Carroll2004}.  Cosmological models in different gravity theories
that predicts a similar expansion rate $H(z)$, can have the different
cosmic growth rate $f(z)$. The $f(z)$ is often parameterized as
$f(z)=\Omega_m^\gamma(z)$ where $\Omega_m(z)$ is the mass density
parameter at a given redshift $z$ and the growth index $\gamma\simeq
0.55 $ in the $\Lambda$CDM model \citep{Linder2005}.  Thus the precise
measurement of the growth rate enables us to investigate the deviation
of gravity from the general relativity.  Recent analysis which focused
on such deviations using weak gravitational lensing data, cosmic
microwave background data, and type Ia supernova data, showed a good
agreement with the pure $\Lambda$CDM model \citep[e.g., see][for the
  latest work]{Daniel2010}.

One of the most promising tools to investigate modified gravity is
redshift-space distortion effects caused by peculiar velocities in
galaxy redshift surveys.  In linear theory and under the
plane-parallel approximation, \citet{Kaiser1987} derived a formula to
relate the observed power spectrum of galaxies $P^{(s)}(k,\mu_{\bf
  k})$ and the true power spectrum of dark matter $P^{(r)}_m(k)$
through
\begin{equation}
  P^{(s)}(k,\mu_{\bf k})=b^2(1+\beta\mu_{\bf
    k}^2)^2P^{(r)}_m(k), \label{eq:kaiser}
\end{equation}
where $(r)$ and $(s)$ respectively denote quantities in real and
redshift space, $\mu_{\bf k}$ is the cosine of the angle between the
line of sight and the wavevector ${\bf k}$, $\beta$ is the linear
redshift distortion parameter related to the growth rate as
$\beta=f/b$, and $b$ is the bias parameter \citep{Kaiser1984}.  Thus
the measurement of the redshift-space distortions allows one to
directly probe deviations from general relativity, although the
determination of the biasing is another important issue.  The Kaiser's
formula (equation (\ref{eq:kaiser})) is modified on small scales
because the nonlinear random velocities smear the clustering along the
line of sight known as the `finger-of-god' effect \citep{Jackson1972}.
However such a nonlinear model still relies on Kaiser's formula on
large scales \citep{Peacock1994}.  For the importance of nonlinearity
on such scales, see \citet{Scoccimarro2004} \citet{Taruya2009},
\citet{Desjacques2010}, and \citet{Jennings2010}

Constraints on $\beta$ have been reported in various surveys
\citep[e.g.,][]{Peacock2001, Zehavi2002, Hawkins2003, Tegmark2004,
  Tegmark2006, Ross2007, Guzzo2008, da Angela2008, Cabre2009}.
\citet{Okumura2008} also showed using a luminous red galaxy sample
from the Sloan Digital Sky Survey (SDSS) that simultaneously analyzing
redshift-space distortions and anisotropy of the baryon acoustic
scales allows one to give a strong constraint on dark energy
equation-of-state, as was theoretically predicted \citep{Hu2003,
  Seo2003, Matsubara2004} and this fact was explicitly emphasized by
\citet{Amendola2005}. \citet{Guzzo2008} considered constraints on $f$
to test the deviation from general relativity using the observations
at different redshifts \citep[see also,][] {Di Porto2008,
  Nesseris2008, Yamamoto2008, Cabre2009}. We note that all these
previous studies have used linear theory prediction of redshift-space
distortions to compare with their measurements on scales presumably
large enough for the linear theory to be valid.  \citet{Nakamura2009}
adopted an alternative approach and constrained the growth factor by
measuring the damping of the baryon acoustic
oscillations. \citet{Reyes2010} gave a strong constraint on modified
gravity theory and confirmed general relativity using the method
proposed by \citet{Zhang2007} which can eliminate the uncertainty of
the galaxy biasing by combining weak gravitational lensing, galaxy
clustering, and redshift-space distortions \citep[for similar
  theoretical attempts, see e.g.,][]{Percival2009, McDonald2009,
  Song2010}.

There are many ongoing and upcoming large galaxy surveys, such as the
SDSS-III's Baryon Oscillation Spectroscopic Survey
\citep[BOSS;][]{Schlegel2009a}, the Fiber Multiobject Spectrograph
\citep[FMOS;][]{Sumiyoshi2009}, the Hobby-Eberly Dark Energy
Experiment \citep[HETDEX;][]{Hill2004}, the WiggleZ
\citep{Glazebrook2007}, the BigBOSS \citep{Schlegel2009b}, and so on.
These observations are expected to enable us to distinguish among
gravity theories with high precision through measurement of
redshift-space distortions as well as that of baryon acoustic
oscillations. However, it is not clear if the accuracy of predicting
redshift-space distortions is better than or comparable to the
precision required from future surveys.  In addition, we do not know
how large the deviation from true cosmology is if any. Precision of
the constraint may depend on galaxy types, such as luminosity and host
halo mass.  There were many attempts to investigate the validity to
use the redshift-space distortions to extract the cosmological
information \citep[e.g.,][]{Hatton1998, Hatton1999, Berlind2001,
  Tinker2006}.  \citet{Tinker2006} found that $\beta$ can be estimated
accurately using linear theory if the finger-of-god effect is removed
perfectly.

In this paper, we present a detailed study on this aspect using a
large set of $N$-body simulations. We measure the power spectra and
correlation functions of dark matter halos. Using them, we estimate
the redshift distortion parameter $\beta$ from the
monopole-to-real-space ratio and the quadrupole-to-monopole ratio,
both of which are related to $\beta$ in linear theory.  Then we
examine whether $\beta$ measured in these ways can give true
cosmological information.  We also investigate the dependence of the
precision of reconstructed $\beta$ on halo mass. Particularly we will
clearly show that the $\beta$ value obtained from the small-halo
clustering does not approach a constant even on large scales as linear
theory predicts.  In addition such small halos are shown to be more
stochastic tracers of the underlying density field than massive halos.
We also discuss in detail on which scale and with which method one can
get the correct $\beta$ or $f$ from the redshift-space distortions.

The paper is organized as follows. In Section \ref{sec:nbody}, we
describe the $N$-body simulations and the halo occupation distribution
model used to populate them with mock galaxies.  The basic two-point
statistics used in our analysis are also presented.  In Section
\ref{sec:theory} we briefly review linear theory of redshift
distortions and how to determine the redshift distortion parameter
$\beta$ from the power spectrum and the correlation function.  Section
\ref{sec:result} is devoted to the analysis of redshift distortion
effects in simulations to determine $\beta$ and the growth rate
$f$. Our conclusions are given in Section \ref{sec:conclusion}.

\section{$N$-body Simulations}\label{sec:nbody}

\subsection{Dark Matter Halo and Galaxy Catalogs}
We use a large set of $N$-body simulations, which is an updated
version of \citet{Jing2007}, to create dark matter halo distribution.
We assume a spatially flat $\Lambda$CDM model with the mass density
parameter $\Omega_m=0.268$, the baryon density parameter
$\Omega_b=0.045$, and the Hubble constant $h=0.71$. Initial conditions
are generated using the matter transfer function by CMBfast code
\citep{Seljak1996} and the density fluctuation amplitude is set to be
$\sigma_8=0.85$.  We employ $1024^3$ particles in 15 cubic boxes of
side $600\himpc$ and 4 of side $1200\himpc$, respectively abbreviated
to L600 and L1200.  We mainly show results obtained from the L1200
boxes, while the L600 boxes are used in order to analyze dark matter
halos with small mass and to see if the L1200 samples have good enough
resolution for the smallest halos.  Simulation parameters are
summarized in Table \ref{tab:simu}.  See \citet{Jing2007} for details
of the simulations.  Dark matter halos are identified at the redshift
$z_{\rm out}$ using the friends-of-friends algorithm with a linking
length equal to 0.2 times the mean particle separation.  All unbound
particles in the FOF halos are further discarded. As shown by
\citet{Jing2007}, it is necessary to eliminate these unbound particles
in order to have a correct measurement of clustering for small halos
of a few tens particles. We use all the halos with more than 12
particles.  Identification of small halos is subtler than that of
massive halos because of the limited number of dark matter particles
which constitute small halos.  As will be shown with L600 and L1200
simulations in Section \ref{sec:statistics} and Section
\ref{sec:result}, however, the clustering of halos can be well
measured to this limit.  We choose $z_{\rm out}\approx 0.28$ because
the luminous red galaxies (LRGs) of the SDSS are at this redshift, but
almost all of our conclusions should not rely on our choice of this
particular redshift.

We consider as a mock galaxy catalog the LRG sample
\citep{Eisenstein2001} in the SDSS \citep{York2000}. In order to
populate the center of the halos with LRGs, we rely on the framework
of the halo occupation distribution \citep[HOD, e.g.,][]{Jing1998,
  Seljak2000, Scoccimarro2001, Berlind2002, Yang2003, Zheng2005},
which describes the relationship between the galaxy and dark matter
density fields.  Galaxies are assigned to the halos using the best fit
HOD parameters for LRGs found by \citet{Seo2008} \citep[see
  also][]{Zheng2005, Zheng2009}.  This method was applied in our
previous work \citep{Okumura2009} and the good agreement with the
observation in clustering has been confirmed.  LRGs are found to
reside in massive halos of typical mass $\sim 2\times 10^{13} -
10^{14}h^{-1}M_{\odot} $. The fraction of satellite LRGs is 6.3\% and
only central LRGs are used for our analysis below.  The peculiar
velocity of their halos is assigned to central LRGs.  Table
\ref{tab:simu2} lists the detail of the representative halo and LRG
catalogs.

\begin{deluxetable}{ccccc}
\tablewidth{0pt} \tablecolumns{5} 
\tablecaption{Simulation parameters} 
\tablehead{\colhead{boxsize}& 
\colhead{particles} & 
\colhead{realizations} & 
\colhead{$m_p (h^{-1}M_\odot)$} & 
\colhead{$z_{\rm out}$}} 
\startdata
 600 & $1024^3$ & 15 & $1.5\times 10^{10}$ & 0.295\\ 
 1200 & $1024^3$ &  4 & $1.2\times 10^{11}$ & 0.274 
 \enddata 
 \tablecomments{
   $m_p$ in column 4 is the particle mass. }
\label{tab:simu}
\end{deluxetable}

\begin{deluxetable}{cccccc}
  \tablecaption{\label{tab:simu2} Properties of simulated halos and galaxies}
  \startdata
  \tableskten\hline\hline \tableskten
  box & $M(h^{-1}M_\odot)$ & $n_p$ & $N_{\rm halo}$ & $b(k)$ & $b(r)$ \\  
  \tableskten\hline \tableskten 
  L600 & $2.2\times 10^{11}$  & $13\leq n_p \leq 18$  & $1.3\times 10^6$ & 0.69 & 0.70 \\
  & $1.7\times 10^{12}$  & $92\leq n_p \leq 136$ & $1.9\times 10^5$ & 0.88 & 0.89 \\
  \tableskten \hline\tableskten
  L1200 & $1.8\times 10^{12}$  & $12\leq n_p \leq 17$  & $1.7\times 10^6$ & 0.89 & 0.88 \\
        & $1.4\times 10^{13}$  & $92\leq n_p \leq 136$ & $2.2\times 10^5$ & 1.30 & 1.30 \\
        & $1.0\times 10^{14}$  & $692\leq n_p \leq 1037$ & $2.2\times 10^4$& 2.28 & 2.31\\
        & LRG  & $12\leq n_p \lesssim 25000$ & $1.4\times 10^5$ & 1.90 & 1.94
  \enddata
  \tablecomments{
    The halo mass $M$ in column 2 shows the central values 
    of each mass bin. Column 3 shows the ranges of the number of particles. 
    $N_{\rm halo}$ is the total number of halos. $b(k)$ and $b(r)$ in columns 4 and 5 
    are the best fit bias parameters in Fourier and configuration space, 
    respectively (see Section \ref{sec:statistics}).}
\end{deluxetable}

\begin{figure*}[bt]
\epsscale{1.}
\plotone{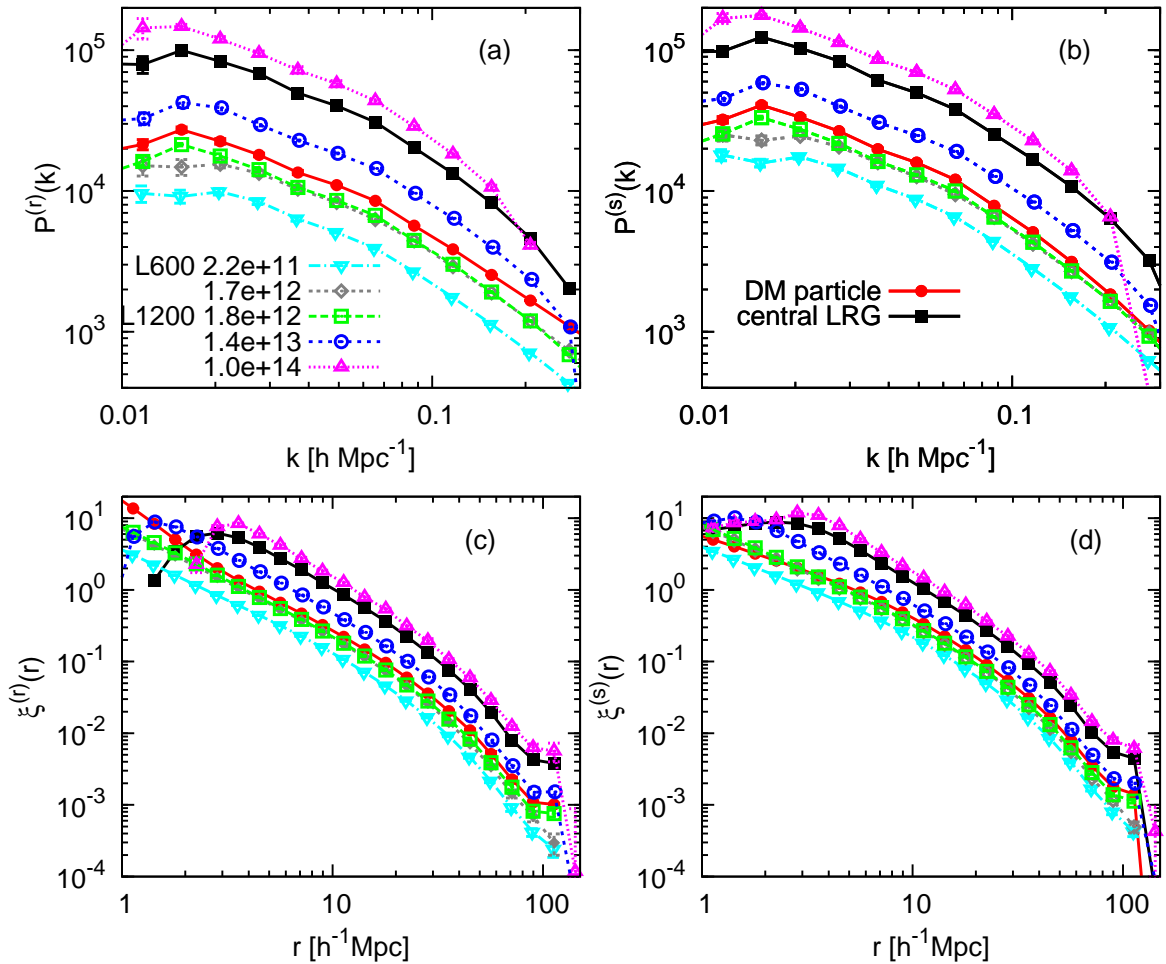}
\caption{Two-point statistics for dark matter halos, LRGs, and dark
  matter particles.  (a) Real-space power spectra $P^{(r)}(k)$.  (b)
  Redshift-space power spectra $P^{(s)}(k)$.  (c) Real-space
  correlation functions $\xi^{(r)}(r)$.  (d) Redshift-space
  correlation functions $\xi^{(s)}(r)$.  The values quoted in the
  figure are the halo mass at the center of mass bin in units of the
  solar mass $[h^{-1}M_\odot]$. Error bars are the standard error of
  the mean.}
\label{fig:xi}
\end{figure*}

\subsection{Two-point Statistics}\label{sec:statistics}

We plot the two-point statistics for dark matter halos, central LRGs,
and dark matter particles measured from our simulation catalogs in
Figure \ref{fig:xi}.  When dark matter halos are analyzed, they are
divided into narrow mass bins as $M_i-\Delta M_i < M_i < M_i+\Delta
M_i$ for $i$th bin where $\Delta M_i=0.2M_i$.  Corresponding ranges of
the number of particles for each halo catalog are listed in Table
\ref{tab:simu2}.  Results of LRGs and dark matter are shown only for
the L1200 samples while those of high and low mass halos are shown for
the L1200 and L600 samples, respectively.  Results are averaged over
all the realizations.  In redshift space, positions of objects are
displaced as a result of the peculiar velocity along the line of
sight. We regard each direction along the three axes of simulation box
as the line of sight and the statistics are averaged over three
projections of all realizations for a total of 45 samples for the L600
simulation and 12 for the L1200 simulation.  The error bars shown in
Figure \ref{fig:xi} are the standard error of the mean.

Figures \ref{fig:xi} (a) and (b) show the power spectra in real space,
$P^{(r)}(k)$, and in redshift space, $P^{(s)}(k)$, respectively. In
both figures clear halo-mass dependence of the clustering amplitude
can be found.  The gray and green lines show the halo power spectra
with the same halo mass in the L600 and L1200 samples,
respectively. It can be easily seen that the agreement of the power
spectra between the two boxes is very good for $k>0.03\hmpci$,
indicating that resolution of a halo with $\ge 12$ particles is enough
for the clustering analysis here. The discrepancy between the two
lines at the smaller $k<0.03\hmpci$ is owing to the large cosmic
variance effect in the L600 sample. Figures \ref{fig:xi} (c) and (d)
show the correlation functions in real space, $\xi^{(r)}(r)$, and in
redshift space, $\xi^{(s)}(r)$, respectively.  The suppression of the
correlation functions for halos and central LRGs is caused by the
finite size of the halos and it is alleviated to some extent in
redshift space due to their random velocities \citep{Jackson1972}.
The baryonic acoustic features, which were clearly detected in our
simulation samples as a single peak at $\sim 100\himpc$, appear as
slight bumps in Figures \ref{fig:xi} (c) and (d) because we use the
binning much broader than the width of the peak.

The bias parameter can be computed both in Fourier space, $b(k)$, and
in configuration space, $b(r)$, through
\begin{equation}
  b(k)=\left( \frac{P^{(r)}(k)}{\pkrm} \right)^{1/2}, \ \ \ 
  b(r)=\left( \frac{\xi^{(r)}(r)}{\xirm} \right)^{1/2}. \label{eq:bias}
\end{equation}
Figure \ref{fig:bias} shows the $b(k)$ and $b(r)$ for dark matter
halos and LRGs. The results are averaged over realizations and the
error bars are the error of the mean.  In both configuration and
Fourier space, the bias parameters for halos and LRGs are found to be
almost constant on sufficiently large scales. We assume the bias to be
constant and search for the best fit value for each sample by
computing the $\chi^2$ statistics.  We compute $\chi^2$ for
$16<r<79\himpc$ in configuration space and for $0.018<k<0.10\hmpci$ in
Fourier space for results from the L1200 simulations. On the other
hand we compute it for $16<r<63\himpc$ in configuration space for the
L600 samples because the large-scale data is not very reliable owing
to the cosmic variance while we still use the same range in Fourier
space.  Figure \ref{fig:bias_m} shows the halo biasing as a function
of the halo mass, $b(M)$.  The error bars show the 95\% confidence
interval. The results between L600 and L1200 are largely overlapped
with each other and we confirm that the systematic error caused by the
different box sizes is negligibly small.  In addition, the bias
parameters obtained in Fourier and configuration space are
consistent. The best fit values obtained here are used for the
theoretical prediction of $\beta$ through Kaiser's formula in Section
\ref{sec:result}.

\section{Linear Theory of Redshift-space Distortions}\label{sec:theory}

There are at least two ways to determine the redshift distortion
parameter $\beta$. They have been well developed both in Fourier space
\citep{Kaiser1987, Cole1994} and in configuration space
\citep{Hamilton1992} under the plane-parallel approximation and
summarized in a review by \citet{Hamilton1998}, who also collected the
observational constraints then available on $\beta$ in various
surveys.  We follow the similar notation with that adopted by
\citet{Tinker2006}.

\subsection{Fourier Space}
For plane-parallel redshift-space distortions, the redshift space
power spectrum can be written as \citep{Kaiser1987}
\begin{equation}
  P^{(s)}(k,\mu_{\bf k})=P_0(k)L_0(\mu_{\bf k})+P_2(k)L_2(\mu_{\bf
    k})+P_4(k)L_4(\mu_{\bf k}),
\end{equation}
where $L_l$ are Legendre Polynomials.  The multipoles of the
redshift-space power spectrum are expressed as
\begin{equation}
  P_l(k)=\frac{2l+1}{2}\int^{+1}_{-1}P^{(s)}(k,\mu_{\bf
  k})L_l(\mu_{\bf k})d\mu_{\bf k}.
\end{equation}
We can derive two useful combinations of them which are directly
related to $\beta$, the ratio of the monopole to the real-space power
spectrum $P^{(0/r)}$ and the quadrupole-to-monopole ratio $P^{(2/0)}$
\citep{Cole1994};
\begin{eqnarray}
  P^{(0/r)}(k) &\equiv& \frac{P_0(k)}{P^{(r)}(k)}  
  = 1+\frac{2}{3}\beta+\frac{1}{5}\beta^2, \label{eq:pk_mono}
  \\
  P^{(2/0)}(k) &\equiv&
  \frac{P_2(k)}{P_0(k)}
  =\frac{\frac{4}{3}\beta+\frac{4}{7}\beta^2}
     {1+\frac{2}{3}\beta+\frac{1}{5}\beta^2}. \label{eq:pk_quad}
\end{eqnarray}
The last equality in the two equations holds only on large scales
where linear theory can be applied.

\begin{figure*}[bt]
\epsscale{1.}  \plotone{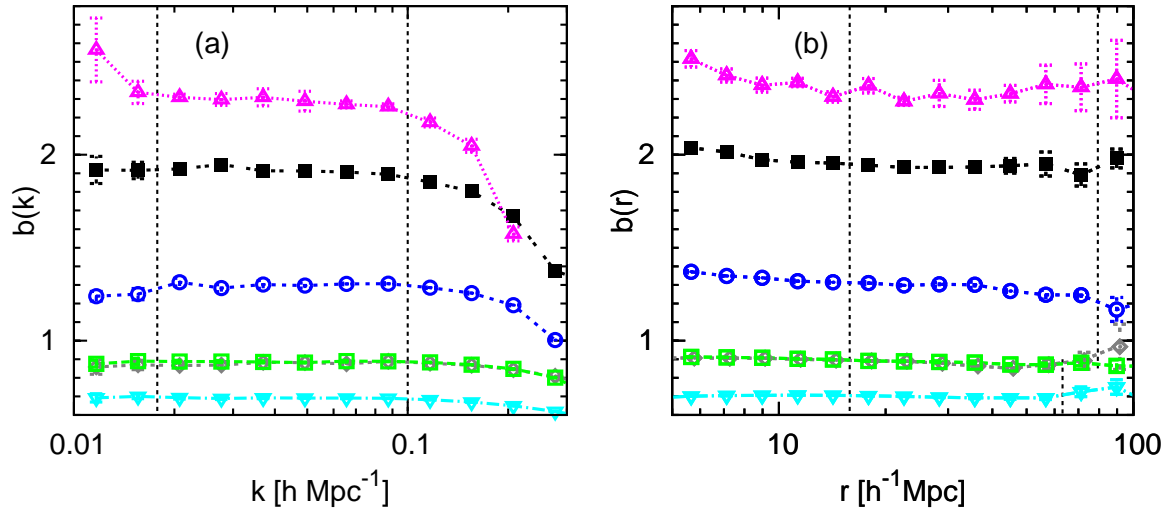}
\caption{Halo bias (a) in Fourier space and (b) in configuration
  space.  The color of each line corresponds to the one with the same
  color in Figure \ref{fig:xi}. The bias for the central LRGs is also
  plotted for comparison. Error bars are the standard error of the
  mean. The region enclosed by the two vertical lines shows the scales
  where we assume the scale-independent bias.  }
\label{fig:bias}
\end{figure*}

\begin{figure}[bt]
\epsscale{1.0}
\plotone{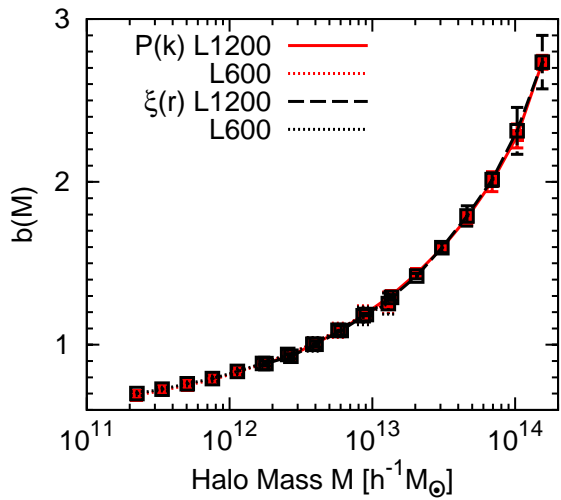}
\caption{Halo bias as a function of the halo mass.  Shown is the best
  fit values for constant fits of $b$ between $15<r<80\himpc$ (L1200)
  and $15<r<60\himpc$ (L600) for $\xi(r)$ and $0.02<k<0.1\hmpci$ for
  $P(k)$. Results from the L1200 samples are shown for $1.4 \times
  10^{12} \leq M \leq 1.9 \times 10^{14}M_\odot$ and those from the
  L600 samples for $1.8\times 10^{11}M \leq 1.6\times 10^{13}M_\odot.$
  The error bars show the 95\% confidence levels. }
\label{fig:bias_m}
\end{figure}

\subsection{Configuration Space}

The redshift-space correlation functions can be expressed similarly to
the power spectra under the plane-parallel approximation as
\begin{equation}
  \xi^{(s)}(r_p,r_\pi)=\xi_0(r)L_0(\mu)+\xi_2(r)L_2(\mu)+\xi_4(r)L_4(\mu),
\end{equation}
where $r_p$ and $r_\pi$ are the separations perpendicular and parallel
to the line of sight and $\mu$ is the cosine of the angle between the
separation vector and the line of sight $\mu=\cos{\theta}=r_\pi/r$.
The multipoles of the redshift-space correlation function are
expressed as
\begin{equation}
  \xi_l(r)=\frac{2l+1}{2}\int^{+1}_{-1}\xi^{(s)}(r_p,r_\pi)L_l(\mu)d\mu.
  \label{eq:xi_multi}
\end{equation}
In linear theory, the ratio of the monopole to the real-space
correlation function and the quadrupole-to-monopole ratio are related
to the redshift distortion parameter $\beta$ on large scales
\citep{Hamilton1992};
\begin{eqnarray}
  \xi^{(0/r)}(r) &\equiv& \frac{\xi_0(r)}{\xi^{(r)}(r)}  
  = 1+\frac{2}{3}\beta+\frac{1}{5}\beta^2, \label{eq:xi_mono}\\
  \xi^{(2/0)}(r) &\equiv&
  \frac{\xi_2(r)}{\xi_0(r)-\bar{\xi_0}(r)}
  =\frac{\frac{4}{3}\beta+\frac{4}{7}\beta^2}
    {1+\frac{2}{3}\beta+\frac{1}{5}\beta^2}, \label{eq:xi_quad}
\end{eqnarray}
where $\bar{\xi_0}(r)=(3/r^3)\int^r_0 \xi_0(r')r'^2dr'$.  When one
wants to constrain the pairwise velocity dispersion of galaxies which
becomes dominant on small scales, the real space correlation function
is convolved with the distribution function of pairwise velocities to
give the redshift space correlation function \citep{Peebles1980},
which is not the purpose of this paper \citep[see,
  e.g.,][]{Peacock2001, Zehavi2002, Hawkins2003, Jing2004,Guzzo2008,
  Cabre2009}. We will briefly discuss the effect of the pairwise
velocities on $\beta$ reconstruction at Section \ref{sec:pvd}.

\section{Results and Discussion}\label{sec:result}

\subsection{$\beta$ Reconstruction}\label{sec:beta}

\begin{figure*}[pbt]
\epsscale{1.0}
\plotone{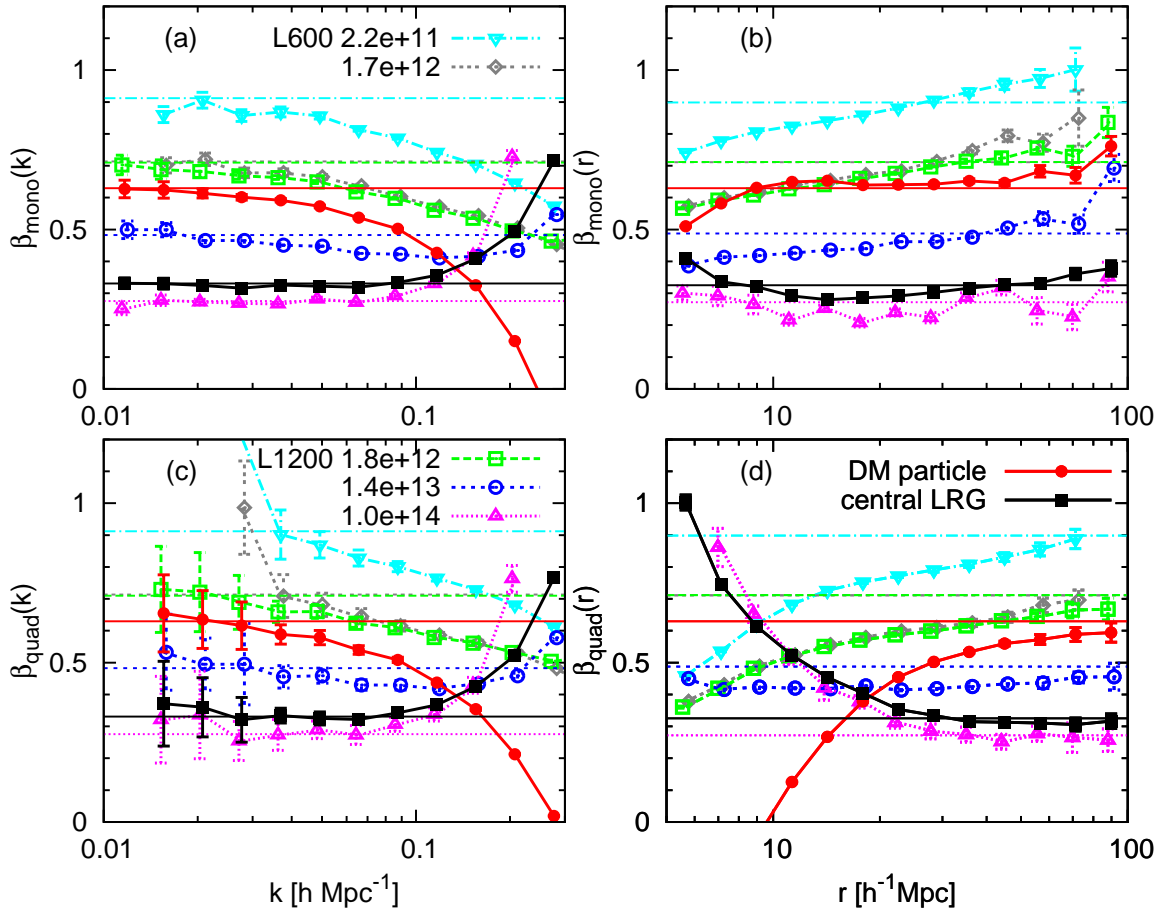}
\caption{Redshift distortion parameter $\beta$ reconstructed from; 
  (a) the monopole-to-real-space ratio of the power spectra; 
  (b) the monopole-to-real-space ratio of the correlation functions;
  (c) the quadrupole-to-monopole ratio of the power spectrum; 
  (d) the quadrupole-to-monopole ratio of the correlation functions. 
  The horizontal lines represent the prediction from linear theory for
  each measurement with the same color and line type, where the best
  fit parameter for the biasing is used for the prediction.  Error
  bars are the standard error of the mean. The diamonds and the open
  circles have been offset in the horizontally positive direction for
  clarity while the open squares and the triangles in the horizontally
  negative direction. }
\label{fig:beta}
\end{figure*}

In Figure \ref{fig:beta} we show the resulting $\beta$ values of dark
matter halos, LRGs, and dark matter reconstructed by the methods
described in Section \ref{sec:theory}. In each panel the horizontal
lines show the large-scale values predicted by general relativity,
$\beta=\Omega_m^{0.55}(z)/b$ \citep{Linder2005}. For the bias
parameters in Fourier and configuration space we use the best fit
values obtained in Figure \ref{fig:bias_m}.  The $\beta$ value of dark
matter is simply equal to the growth rate $f$ because $b=1$.  We can
see the agreement of the $\beta$ values obtained from the L600 and
L1200 samples with the same halo mass, thus the different number of
particles, indicating that resolution of a halo with 12 particles is
accurate enough for $\beta$ reconstruction. The discrepancy between
the two on large scales is again owing to the cosmic variance in the
L600 sample.

Figure \ref{fig:beta} (a) shows $\beta$ as a function of $k$ measured
from the ratio of the monopole to the real-space power spectrum
$P^{(0/r)}=P_0/P^{(r)}$. Small-scale values obtained from the dark
matter particles and small-mass halos are suppressed by the random
peculiar velocities, while those from the LRGs and massive halos go up
on such scales as a result of the finite size of halos, as described
in Section \ref{sec:statistics}.  One can see that, for the most
massive halos ($M\sim 10^{14}h^{-1}M_\odot$) and LRGs, the ratio
$P^{(0/r)}$ reconstructs the $\beta$ values predicted by linear theory
at $k<0.08\hmpci$.  This means that linear theory is accurate enough
to predict and constrain $\beta$ using such massive halos. On the
other hand, the smaller halos we focus on, the more prominent scale
dependence of $\beta$ we find even on large scales.  This behavior of
$\beta$ from the smaller mass halos is consistent with that obtained
by \citet{Tinker2006}.  They used mock galaxies assigned to the halos
by the HOD parameters applied to the SDSS MAIN galaxies
\citep{Zehavi2005} which preferentially reside in halos with small
mass. Thus the behavior seen by \citet{Tinker2006} is found to be
caused by the contribution from smaller halos.  Note that here we used
the real-space power spectrum measured from the simulations. It is not
a direct observable and translation of the power spectrum from
redshift space to real space usually draws the additional error.

Figure \ref{fig:beta} (b) is the same as Figure \ref{fig:beta} (a),
but the $\beta$ values are measured as a function of separation $r$
computed from the ratio of the redshift-space and real-space
correlation functions $\xi^{(0/r)}=\xi_0/\xi^{(r)}$.  The slight
difference of the linear theory prediction between Fourier and
configuration space is due to the difference of the best fit
parameters for the biasing seen in Figure \ref{fig:bias_m}.  The scale
dependence of the reconstructed $\beta$ found by using $\xi^{(0/r)}$
is more prominent than that by using $P^{(0/r)}$.  Even the result
obtained from the LRGs is monotonically increasing, intersects with
the prediction from linear theory, and does not draw closer to a
constant on all scales probed.  A similar behavior has also been found
for dark matter by \citet{Cabre2009} up to $40\himpc$ (see also the
red line in Figure 4b), but the tendency is much more significant for
dark matter halos, even for those with $b\sim 1$.  Constraints on
$\beta$ are usually given under the assumption of one constant
parameter over a scale range for which the $\chi^2$ statistics is
computed. However, according to Figure \ref{fig:beta} (b), it could be
a coincidence that one gets the true value of $\beta$ as a best fit
parameter.  Thus one should be cautious when the deviation from
general relativity is investigated through the measurement of $\beta$
from the ratio $\xi_0/\xi^{(r)}$.

In measuring $\beta$ from the quadrupole-to-monopole ratio in Fourier
space $P^{(2/0)}=P_2/P_0$, one needs to measure $P^{(s)}(k,\mu)$ in
finite bins, usually constant separations in $\mu$, and numerically
integrate it along $\mu$ direction. Hence the finite bin size may
cause a systematic error in measurement of $\beta$. Using linear
theory, we test the accuracy of the integration between constant $\mu$
and constant $\theta=\cos^{-1}{\mu}$ binnings. We found constant $\mu$
binning underestimates $\beta$ by 2.5\% while constant $\theta$
binning overestimates by 1.3\% for 10 bins.  We thus adopt the
constant $\theta$ binning and take the number of bin to be 10 between
$0\leq\theta\leq90^\circ$.  Figure \ref{fig:beta} (c) shows $\beta$
measured from the quadrupole-to-monopole moments $P^{(2/0)}$. This
quantity can be directly measured in observation. We put artificial
large-scale cuts in $\beta$ values measured from the L600 samples
because they have limited number of modes and thus are strongly
affected by the cosmic variance. We can see this by the difference
between the green and gray lines because they have similar halo
mass. On the other hand, we did not use such strong scale cuts in
Figure \ref{fig:beta} (a) because the effect of the cosmic variance
can be eliminated to some extent by taking the ratio of two power
spectra \citep{McDonald2009}.  In Fourier space, the behavior of
$\beta$ measured from $P^{(2/0)}$ is almost the same as that from
$P^{(0/r)}$ except for the magnitude of the error bars owing to the
cosmic variance.

Finally, we show $\beta$ values measured from the
quadrupole-to-monopole ratio in configuration space
$\xi^{(2/0)}=\xi_2/\xi_0$ in Figure \ref{fig:beta} (d).  In measuring
the quadrupole moment by equation (\ref{eq:xi_multi}) we adopt the
binning on a polar grid of logarithmic spacing in $r$ and linear
spacing in angle, then numerically integrate the correlation function
at each $r$, as was done by \citet{Tinker2006}. On small scales, the
behavior of $\beta$ thus determined is complicated. For small halos,
the results give lower $\beta$ values than those from linear theory
prediction, which may be caused by the random peculiar velocities and
thus might be correctable to some extent \citep[see][]{Hawkins2003,
  Tinker2006}. However, note that peculiar velocities predicted from a
simple halo model \citep{Yang2003} have very different luminosity
dependence with those from observations \citep[see also Slosar et
  al. 2006]{Jing2004, Li2006}. For the massive halos, the $\beta$
values on small scales are larger than the linear theory prediction,
which might indicate that these halos are approaching with each
other. Interestingly, for the most massive halos ($1.0\times 10^{14}
h^{-1}M_\odot$) and the LRG sample, we can simply use linear theory
and use the data points on scales larger than $\sim 25\himpc$ in order
to constrain the growth rate.  For galaxies within the halos with the
mass $\sim 1.4\times 10^{13}h^{-1}M_\odot$, the $\beta$ value is
coincidently a constant, but it is lower than the linear theory
prediction. If we use a population of such galaxies, we would
underestimate the growth factor.  Lastly for galaxies within less
massive halos, the reconstructed $\beta$ becomes a scale-dependent
function, and one has to be extremely careful in extracting the linear
growth factor from a measurement of redshift-space distortions of such
galaxies.

\begin{figure}[t]
\epsscale{1.00}
\plotone{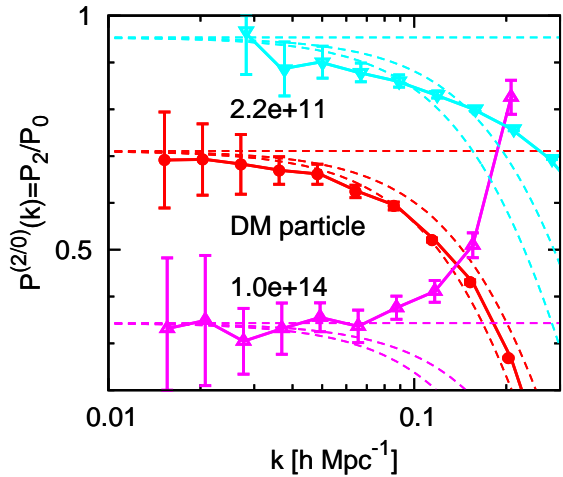}
\caption{Quadrupole-to-monopole ratio in Fourier space for dark matter
  particles (red) and for massive (magenta) and small (light blue)
  halos.  The horizontal lines are the linear theory prediction. The
  upper and lower dashed curves in red are the non-linear prediction
  with the pairwise velocity dispersion $\sigma_v=500$ and $600 {\rm
    km/s}$, respectively, while those for halos are with
  $\sigma_v=400$ and $500 {\rm km/s}$.  }
\label{fig:p20}
\end{figure}

\subsection{Pairwise Velocity Dispersion}\label{sec:pvd}
In order to see if random peculiar velocities can cause the deviation
from the linear theory prediction, we consider a simple Exponential
model for the pairwise velocity dispersion (PVD) in configuration
space which results in a Lorentz damping factor in Fourier space,
\begin{equation}
	G(k,\mu_{\bf k},\sigma_v)=(1+k^2\mu^2_{\bf k}\sigma^2_v/2)^{-1}.
\end{equation}
The power spectrum of this dispersion model is expressed as Equation
(\ref{eq:kaiser}) multiplied by $G(k,\mu_{\bf k},\sigma_v)$.  Although
more accurate models have been developed by many authors, this simple
model is useful enough for our purpose.  In Figure \ref{fig:p20}, we
show the predictions for the quadrupole-to-monopole ratio in Fourier
space $P^{(2/0)}$ from the dispersion model.  In order to avoid making
the figure unclear, we show the results for only dark matter
particles, the most massive halos and the smallest halos.  The overall
shape of $P^{(2/0)}$ for dark matter is well explained by the
dispersion model with $\sigma_v\sim 600$ km/s, which has already been
found using more accurate models \citep[e.g., ][]{Taruya2010}.
However, the results for dark matter halos are much more complicated.
Here we adopt $\sigma_v\sim 450$ km/s for the halos according to
\citep{Hamana2003}.  For small halos, not only the deviation of the
measurement from the linear theory prediction but also its scale
dependence cannot be corrected by the model. On the other hand, the
behavior of $\beta$ reconstructed for massive halos is opposite to the
dispersion model.  These complex results are somewhat expected because
the halo peculiar velocities change very mildly with their mass
\citep{Hamana2003}. Thus the difference of the scale dependences of
$\beta$ among small and large halos cannot be simultaneously explained
with such analytical models for the PVD. We note that the model of PVD
was adopted by \citet{Guzzo2008} to correct for the nonlinear effect
when they measure the $\beta$ at redshift 0.8.

\begin{figure*}[tb]
\epsscale{1.00}
\plottwo{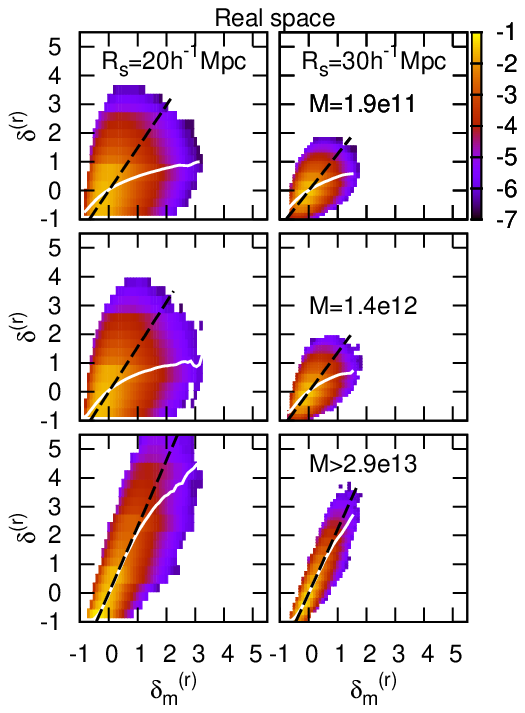}{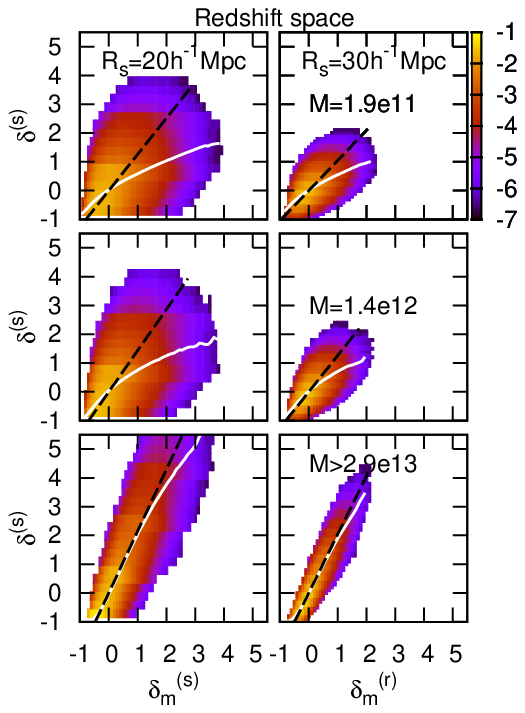}
\caption{Joint probability distributions of overdensity fields for
  dark halos with dark matter overdensity in real space ({\it left
    panels}) and in redshift space ({\it right panels}) smoothed over
  $R_s=20\himpc$ and $R_s=30\himpc$. Upper, middle, and lower panels
  in each set show the results for halos with $M=1.9\times 10^{11}$,
  $M=1.4\times 10^{12}$, and $M>1.9\times 10^{13}$, respectively.  The
  values in the color bars show the logarithm of the probability to
  base 10.  The solid lines are the conditional mean
  $\bar{\delta}(\delta_m)$.  The dashed lines show the linear bias
  relation $\delta(\delta_m)=b_{{\rm var}}\delta_m$.  }
\label{fig:stochasticity} 
\end{figure*}

\begin{figure*}[tb]
\epsscale{1.00}
\plottwo{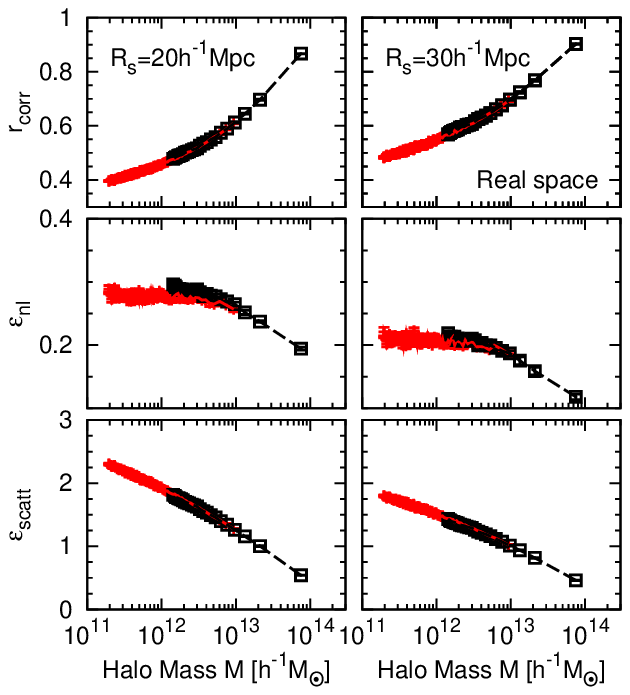}{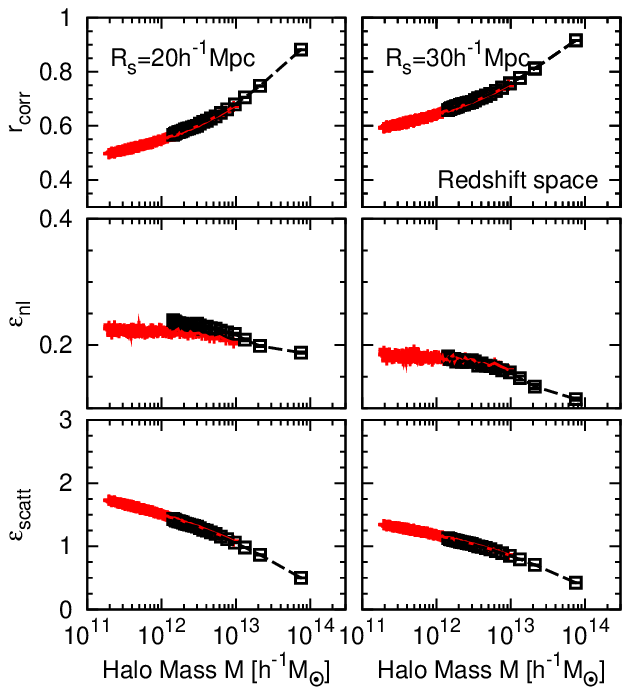}
\caption{Dependence of $r_{{\rm corr}}$, $\epsilon_{{\rm nl}}$, and
  $\epsilon_{{\rm scatt}}$ in real space ({\it left panels}) and in
  redshift space ({\it right panels}) smoothed over $R_s=20\himpc$ and
  $R_s=30\himpc$.  The red points show the results obtained from the
  L600 samples while the black points the L1200 samples. Error bars
  are the standard error of the mean. }
\label{fig:stochastic_para} 
\end{figure*}

\subsection{Nonlinear Stochastic Biasing}
In the derivation of Kaiser's effect, the linear bias relation between
objects considered (galaxies or halos) and dark matter was
assumed. However, there could be considerable stochasticity and
nonlinearity in the bias relation between dark halos (or galaxies) and
dark matter \citep{Dekel1999}.  We follow the formalism proposed by
\citet{Taruya2000} and applied to simulation data by
\citet{Yoshikawa2001}.  We briefly summarize some parameters defined
by \citet{Taruya2000} to quantify the nonlinearity and stochasticity
of the halo bias. First, the density fields of dark matter and dark
halos are evaluated as $\delta_m({\bf x},R_S)$ and $\delta({\bf
  x},R_S)$, respectively, smoothed over the top-hat window radius
$R_s$.  The bias parameter and the correlation coefficient are
respectively defined by
\begin{equation}
b_{{\rm var}}\equiv \sqrt{\frac{\left\langle \delta^2
    \right\rangle}{\left\langle \delta_m^2 \right\rangle}},
\ \ \ r_{{\rm corr}}\equiv \frac{\left\langle
  \delta\delta_m\right\rangle}{\sqrt{\left\langle \delta^2
    \right\rangle\left\langle \delta_m^2 \right\rangle}}.
\end{equation}
Note that the bias parameters defined in Section \ref{sec:statistics}
are from the two-point statics, while the bias $b_{{\rm var}}$ defined
here is from one-point statistics.  In order to quantify the nonlinear
and stochastic nature of the biasing separately, two more useful
parameters are introduced. For this purpose, let us define the
conditional mean of $\delta$ for a given $\delta_m$,
\begin{equation}
	\bar{\delta}(\delta_m)=\int \delta \ P(\delta |
        \delta_m)d\delta,
\end{equation}
where $P(\delta | \delta_m)$ is the conditional probability
distribution function.  Then the nonlinearity of the biasing is
quantified by
\begin{equation}
\epsilon_{{\rm nl}}\equiv \frac{\left\langle\delta^2_m\right\rangle \left\langle\bar{\delta}^2\right\rangle}{\left\langle \bar{\delta}\delta_m\right\rangle^2}-1,
\end{equation}
which vanished only when the biasing is linear. Similarly, the stochasticity of the biasing is characterized by 
\begin{equation}
\epsilon_{{\rm scatt}}\equiv  \frac{\left\langle\delta^2_m\right\rangle \left\langle (\delta - \bar{\delta})^2\right\rangle}{\left\langle \bar{\delta}\delta_m\right\rangle^2},
\end{equation}
which vanishes for the deterministic bias where $\delta=\bar{\delta}(\delta_m)$.

Following the same procedure of \citet{Yoshikawa2001}, we compute the
parameters defined above for our halo catalogs in real and redshift
space.  In order to minimize the Poisson noise effect and fairly
compare the results of different mass, only in this subsection we keep
each subsample of a given halo mass to have the same number density,
$1.16\times 10^{-4}(h^{-1}{\rm Mpc})^{-3}$.  This density corresponds
to $N_{{\rm halo}}\approx 2.5\times 10^4$ for L600 samples and
$N_{{\rm halo}}\approx 2.0\times 10^5$ for L1200 samples.  We adopt
the smoothing scales $R_s=20$ and 30\himpc.  Many pairs of the values
$[\delta({\bf x}), \delta_m({\bf x})]$ are obtained for randomly
selected points ${\bf x}$ in the simulation box.

In Figure \ref{fig:stochasticity}, we show the joint distribution of
$\delta$ with $\delta_m$ in real space (left) and in redshift space
(right). From the top to bottom, the results obtained from the
smallest mass bin of the L600 samples ($M=1.9\times
10^{11}h^{-1}M_\odot$), the smallest mass bin of the L1200 samples
($M=1.4\times 10^{12}h^{-1}M_\odot$), and the largest mass bin of the
L1200 samples ($M>2.9\times 10^{13}h^{-1}M_\odot$) are plotted.  In
each panel we also plot the conditional mean relation
$\bar{\delta}(\delta_m)$ as the solid line and the linear bias
relation $\delta=b_{{\rm var}}\delta_m$ as the dashed line, both of
which are obtained from our simulations.  Here we focus on halo mass
dependence of the nonlinear stochastic biasing.  The deviation of
$\bar{\delta}$ from the linear bias is caused by the nonlinear
stochastic bias as well as the halo exclusion effect \citep[][see also
  Smith et al. 2007]{Yoshikawa2001}.  The halo exclusion effect is
alleviated in redshift space due to the random velocities of halos, as
we have seen in Section \ref{sec:statistics}.  Despite the fact that
this exclusion effect is more significant for larger thus more massive
halos, the deviation from the linear relation for such halos is much
smaller.  This indicates that more massive halos have the smaller
nonlinearity and stochasticity, and the latter was also found by
\citet{Hamaus2010} using a complementary statistics in real space.

In order to see these effects more quantitatively, we show $r_{{\rm
    corr}}$, $\epsilon_{{\rm nl}}$, and $\epsilon_{{\rm scatt}}$ as
functions of halo mass in Figure \ref{fig:stochastic_para} from the
top to bottom.  The results for the parameter $\epsilon_{{\rm nl}}$
show that the nonlinearity of the halo biasing is smallest for the
most massive halos.  It increases for smaller halos and gets close to
a constant.  Similarly, the stochasticity parameter $\epsilon_{{\rm
    scatt}}$ has the minimum value for the most massive halos.  At a
whole mass range probed, however, the stochastic bias monotonically
increases toward the lowest mass.  Finally the stochasticity of the
smallest halos becomes five times larger than that of the most massive
halos.  Thus both the nonlinearity and stochasticity of the halo
biasing, particularly the latter, is likely one of the cause for the
systematic deviation of $\beta$ values from the theoretical
prediction.

\subsection{Halo Mass Dependence of Growth Rate Constraints}\label{sec:growth}

\begin{figure*}[bt]
\epsscale{1.00} \plotone{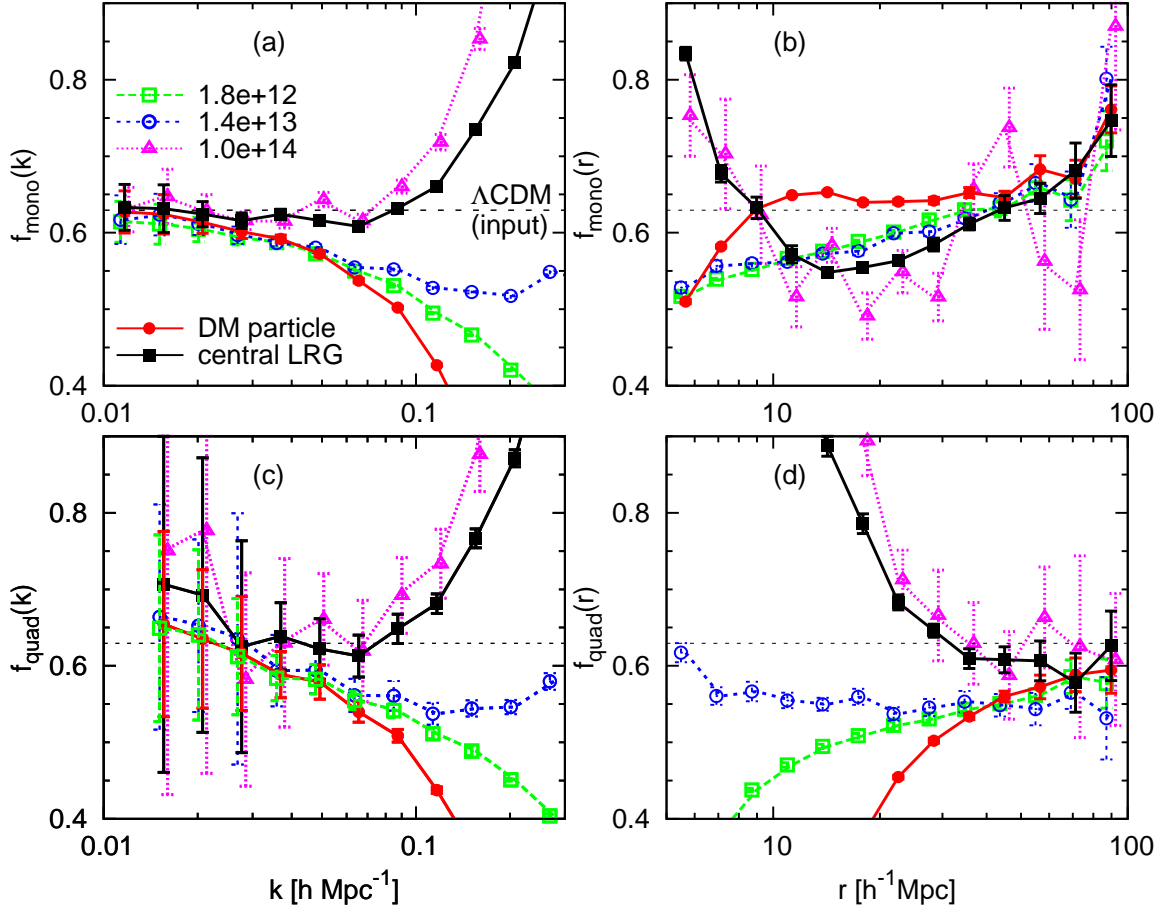}
\caption{Same as Figure \ref{fig:beta} but for growth rate $f$,
  reconstructed from; 
  (a) the monopole-to-real-space ratio of the power spectra; 
  (b) the monopole-to-real-space ratio of the correlation functions;
  (c) the quadrupole-to-monopole ratio of the power spectrum; 
  (d) the quadrupole-to-monopole ratio of the correlation functions. 
  The horizontal line shows the linear theory prediction,
  $\Omega_m^{0.55}(z)$. Error bars are the standard error of the
  mean. The open squares and the open circles have been offset in the
  horizontally negative direction for clarity while the triangles in
  the horizontally positive direction. }
\label{fig:f}
\end{figure*}

Because we also calculated the power spectra and correlation functions
for dark matter particles in the same samples as those for dark matter
halos and LRGs, we can directly reconstruct the growth rate through
\begin{eqnarray}
  f(r)=b(r)\beta(r)&=&\left[ \frac{\xi^{(r)}(r)}{\xirm} \right]^{1/2}\beta(r), \\
  f(k)=b(k)\beta(k)&=&\left[ \frac{P^{(r)}(k)}{\pkrm} \right]^{1/2}\beta(k).
\end{eqnarray}
The difference between the use of $\beta$ and $f$ is just whether the
bias $b$ is used as prediction or measurement. When $f$ is used,
however, we can take into account the slight scale dependence of the
bias seen in Figure \ref{fig:bias}.  Figures \ref{fig:f} show the
growth rate measured from the four methods described above.  The
horizontal line shows the input $\Lambda$CDM model predicted from
general relativity $f=\Omega_m^{0.55}(z)$ \citep{Linder2005}.  We can
also see that the linear redshift distortions for the LRGs reconstruct
the true value of $f$ well, except for that from $\xi^{(0/r)}$.

\begin{figure}[t]
\epsscale{1.00}
\plotone{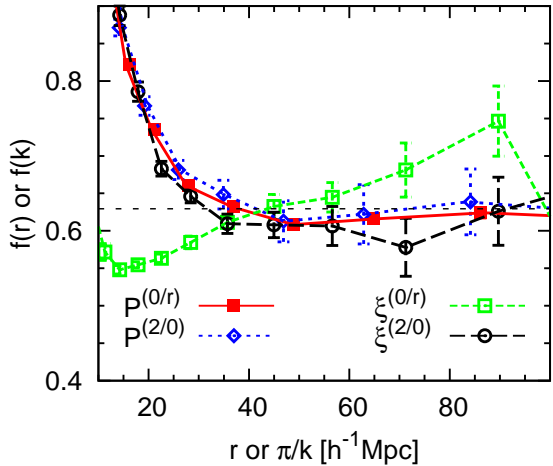}
\caption{Comparison of the growth rate measurement from LRGs among the
  different methods. The horizontal axis shows separation $r$ for
  configuration space measurement while $\pi/k$ for Fourier space
  measurement.  The diamonds/filled squares have been offset in the
  horizontally negative/positive direction for clarity.}
\label{fig:f_comp}
\end{figure}

Here let us discuss which method can be used to obtain the growth
factor $f$ better.  Figure \ref{fig:f_comp} shows the comparison among
the 4 methods for the measurement of $f$ from the LRG clustering.  As
we have seen above, we find strong scale dependence of the growth
factor obtained from $\xi^{(0/r)}$. On the other hand, both of the
methods in Fourier space, $P^{(0/r)}$ and $P^{(2/0)}$, and the
estimator $\xi^{(2/0)}$ in configuration space, can give good
estimation of $\beta$ in linear theory on scales $\lambda \equiv \pi/k
> 30\himpc$ or $r > 30\himpc$.

\section{Conclusion}\label{sec:conclusion}
We have investigated how accurately the redshift-space distortions can
be used to measure the linear growth factor $f$. The growth factor is
a powerful observable targeted by future large redshift surveys to
probe dark energy and to distinguish among different gravity theories.
For this purpose, we constructed a large set of $N$-body simulations,
dividing each dark matter halo catalog into the subsamples with narrow
mass ranges.  As an example of a galaxy sample, mock SDSS LRG catalogs
were constructed by applying the HOD modeling to the simulated halos.
Then we have measured the two-point statistics, power spectra and
correlation functions, for dark matter halos and LRGs. The dark matter
halos were analyzed as a function of halo mass in order to see
dependence of the $\beta$ measurement on the halo mass.

We have determined the $\beta$ values as a function of halo mass and
scale using four methods. First, we found that $\beta$ reconstructed
from the ratio of the monopole to the real-space power spectra
$P^{(0/r)}=P^{(s)}/P^{(r)}$ (equation (\ref{eq:pk_mono}))
asymptotically approaches the true value.  In particular for the
massive halos and LRGs, the prediction from linear theory known as
Kaiser's formula is applicable to give a correct constraint on the
growth rate. However, for less massive halos, the ratio approaches the
true value only at a very large scale $k<0.02\hmpci$.  Second, $\beta$
reconstructed from the ratio of the monopole to the real-space
correlation function $\xi^{(0/r)}=\xi^{(s)}/\xi^{(r)}$ (equation
(\ref{eq:xi_mono})) approaches neither the true value nor a constant
even on large scales. This statement is valid especially for small
halos with the bias parameter $b\leq 1$. Because the growth rate is
assumed to be a constant when modified gravity theories are tested,
the ratio $\xi^{(s)}/\xi^{(r)}$ cannot be used in a simple way for
this purpose.  Third, the quadrupole-to-monopole ratio in Fourier
space $P^{(2/0)}=P_2/P_0$ (equation (\ref{eq:pk_quad})) gives almost
the same value of $\beta$ as $P^{(0/r)}$ but with larger error bars as
expected.  Finally, we found that when the quadrupole-to-monopole
ratio in configuration space $\xi^{(2/0)}=\xi_2/\xi_0$ (equation
(\ref{eq:xi_quad})) is used, a similar conclusion is reached to that
of $P^{(2/0)}$ when $r=\lambda=\pi/k$ is adopted.

For small halos with $b\leq 1.3$, the reconstructed $\beta$ values do
not approach a constant in most of measurable regions, particularly
those from $\xi^{(0/r)}$ in the configuration space.  No method can
provide a reliable estimator for the determination of the growth
factor from the clustering of such small halos on the large range of
scales probed.  Using the halo catalogs with different box sizes, we
confirmed that such a behavior is not caused by the resolution effect
of small dark matter halos.  While the scale dependence changes with
the halo mass, the peculiar velocity of halos does not change much
with the mass \citep{Hamana2003}.  Using the simple dispersion model,
we demonstrated that the different scale dependence of $\beta$ among
small and large halos cannot be simultaneously explained. Also there
are two types of velocity biases which affect the redshift distortion;
the dynamical bias caused by dynamical friction and the spatial bias
caused by the difference between the distribution of halos and that of
dark matter. There is no dynamical velocity bias because the halo
velocities are determined from the mean velocities of dark matter
within the halos in our analysis. The spatial velocity bias should
exist, which is coupled with the nonlinear stochastic bias discussed
in the text.

On the other hand, it is known that the clustering of small dark
matter halos depends not only on their mass but also on their assembly
history, so called the assembly bias
\citep[e.g.,][]{Gao2005}. \citet{Wang2007} showed that old and
low-mass halos that are preferentially associated with a high density
field are more strongly clustered than young halos with the same mass,
and consequently have higher velocities.  Besides, the stochasticity
between halos and dark matter is \citep{Dekel1999} can be a source of
the systematic errors in the $\beta$ reconstruction.  Using a method
introduced by \citet{Taruya2000} and applied to simulation data by
\citet{Yoshikawa2001}, we have found that both the nonlinearity and
the stochasticity of small halos become larger than massive
halos. Particularly the stochastic bias monotonically increases as the
mass of halos decreases, as was found in real space by
\citet{Hamaus2010} using the two-point statistics.  Thus the strong
scale dependence of $\beta$ for low mass halos could be caused by the
assembly and/or nonlinear stochastic bias. However fortunately, the
scale dependence of the measured $\beta$ weakens with the increase of
halo mass. For massive halos with $b>1.5$, the measured $\beta$
approaches the constant predicted by Kaiser's formula on scales
$k<0.08 \hmpci$ or $r>30 \himpc$.

Because the analysis of redshift-space distortions is powerful to
investigate not only properties of dark energy but also modified
gravity theories, it will keep on playing a key role in ongoing and
upcoming large redshift surveys, such as BOSS, FMOS, HETDEX, WiggleZ,
and BigBOSS. Galaxies targeted by the BOSS survey are luminous red
galaxies, which reside in massive halos. In this work we demonstrate
the $\beta$ value can be well reconstructed with a redshift distortion
analysis of LRGs. On the other hand, one of the samples targeted by
the BigBOSS, for example, is that of emission-line galaxies, which
reside in halos with a broad range of halo mass. One needs to be
careful in using such a sample to constrain the growth rate from the
redshift distortion, because it can be a scale-dependent
function. While one might be able to obtain a result consistent (or
inconsistent) with the prediction from general relativity, it could be
just a coincidence after the scale-dependent growth rate is averaged
over some separation or wavenumber ranges. We will use semi-analytical
modeling or a halo occupation model to investigate this issue in
future work.

Recently, an interesting method was proposed by \citet{Seljak2009} to
suppress the shot noise in power spectrum measurement.  They
considered an optimal weighting function $f(M)$ in measuring the
galaxy overdensity, where they give higher weights on higher mass
halos.  Compared with our results presented here, such a mass
weighting scheme is useful not only for suppressing the shot noise but
also obtaining the true value of the growth rate. This scheme can be
naturally incorporated into our method and such a study will be
presented in our future paper.  Another theoretical improvement to be
applied for observation is evading the cosmic variance limit, which is
one of the most important tasks for precise measurement of the
redshift-space distortions, as we have already seen above.
\citet{McDonald2009} showed using multiple tracers of density with
different biases suppresses the noise for measurement of $\beta$ on
large scales dramatically compared to the traditional single tracer
method \citep[see also][]{White2009, Gil-Marin2010}. But the different
scale-dependent properties of $\beta$ for different halo masses found
in Figure 4 imply that the real situation might be more complex, and
realistic models of galaxies must be adopted to investigate if the
method of multiple tracers works.

\acknowledgments 

We would like to thank Uro\v{s} Seljak and Vincent Desjacques for
fruitful discussions and Issa Kayo for useful comments.  We also thank
an anonymous referee for many useful suggestions.  This research in
Seoul was supported by WCU (World Class University) program through
the National Research Foundation of Korea funded by the Ministry of
Education, Science and Technology (R32-2009-000-10130-0), and the
research in Shanghai was supported by NSFC (10821302, 10878001), by
the Knowledge Innovation Program of CAS (No. KJCX2-YW-T05), by 973
Program (No. 2007CB815402), and by the CAS/SAFEA International
Partnership Program for Creative Research Teams (KJCX2-YW-T23).

\end{document}